\documentclass[pdflatex,sn-mathphys-num]{sn-jnl}
\usepackage{lineno}

\usepackage{graphicx}%
\usepackage{multirow}%
\usepackage{amsmath,amssymb,amsfonts}%
\usepackage{amsthm}%
\usepackage{mathrsfs}%
\usepackage[title]{appendix}%
\usepackage{xcolor}%
\usepackage{textcomp}%
\usepackage{manyfoot}%
\usepackage{booktabs}%
\usepackage{algorithm}%
\usepackage{algorithmicx}%
\usepackage{algpseudocode}%
\usepackage{listings}%

\usepackage{siunitx}

\newcommand{\angstrom}{\text{\normalfont\AA}}

\begin{document}

\title{\textbf{Atomistic prediction of magnetic behaviour across chemistry, structure and temperature}}

\author[1,2]{\fnm{Ruosong} \sur{Shi}}
\equalcont{These authors contributed equally to this work.}

\author[1]{\fnm{Chengbing} \sur{Chen}}
\equalcont{These authors contributed equally to this work.}

\author[3,4]{\fnm{Bicheng} \sur{Liu}}
\equalcont{These authors contributed equally to this work.}

\author[5]{\fnm{Daye} \sur{Zheng}}

\author[5,6,7]{\fnm{Duo} \sur{Zhang}}

\author[8]{\fnm{Zhengtao} \sur{Huang}}

\author[9]{\fnm{Teng} \sur{Yang}}

\author[10]{\fnm{Zefeng} \sur{Cai}}

\author[11]{\fnm{Xingliang} \sur{Peng}}

\author[6]{\fnm{Xinzijian} \sur{Liu}}

\author[12]{\fnm{Shenshen} \sur{Yan}}

\author[12]{\fnm{Xueqing} \sur{Zhao}}

\author[15]{\fnm{Jianbo} \sur{Wang}}

\author[4]{\fnm{Bin} \sur{Liu}}

\author*[13,14]{\fnm{Han} \sur{Wang}}\email{wang\_han@iapcm.ac.cn}

\author*[2]{\fnm{Ke} \sur{Wang}}\email{wang\-ke@tsinghua.edu.cn}

\author*[3,4]{\fnm{Liuliu} \sur{Han}}\email{hanll@szlab.ac.cn}

\author*[1]{\fnm{Ben} \sur{Xu}}\email{bxu@gscaep.ac.cn}

\affil[1]{\orgname{Graduate School of China Academy of Engineering Physics}, \linebreak \orgaddress{\city{Beijing}, \postcode{100193}, \country{P. R. China}}}

\affil[2]{\orgdiv{State Key Laboratory of New Ceramics and Fine Processing, School of Materials Science and Engineering}, \orgname{Tsinghua University}, \linebreak \orgaddress{\city{Beijing}, \postcode{100084}, \country{P. R. China}}}

\affil[3]{\orgdiv{} \orgname{Suzhou National Laboratory}, \orgaddress{\city{Suzhou}, \postcode{215123}, \state{} \country{P. R. China}}}

\affil[4]{\orgdiv{State Key Laboratory of Powder Metallurgy}, \orgname{Central South University}, \orgaddress{\city{Changsha}, \postcode{410083}, \country{P. R. China}}}

\affil[5]{\orgname{AI for Science Institute}, \orgaddress{\city{Beijing}, \postcode{100080}, \country{P. R. China}}}

\affil[6]{\orgname{DP Technology}, \orgaddress{\city{Beijing}, \postcode{100080}, \country{P. R. China}}}

\affil[7]{\orgdiv{Academy for Advanced Interdisciplinary Studies}, \linebreak \orgname{Peking University}, \orgaddress{\city{Beijing}, \postcode{100871}, \country{P. R. China}}}

\affil[8]{\orgdiv{Center for Structural Materials, Department of Mechanical Engineering},  \orgname{The University of Hong Kong}, \orgaddress{\city{Hong Kong}, \postcode{999077}, \country{P. R. China}}}

\affil[9]{\orgdiv{European Theoretical Spectroscopy Facility, Institute of Condensed Matter and Nanosciences}, \orgname{Université catholique de Louvain}, \orgaddress{\city{Louvain-la-Neuve}, \postcode{1348}, \country{Belgium}}}

\affil[10]{\orgdiv{Department of Materials Science and Engineering}, \linebreak \orgname{Carnegie Mellon University}, \orgaddress{\city{Pittsburgh}, \postcode{15213}, \state{PA}, \country{USA}}}

\affil[11]{\orgname{SINOPEC (Beijing) Research Institute of Chemical Industry Co., Ltd.}, \linebreak \orgaddress{\city{Beijing}, \postcode{100013}, \country{P. R. China}}}

\affil[12]{\orgname{MetaX Integrated Circuits (Shanghai) Co., Ltd.}, \linebreak \orgaddress{\city{Shanghai}, \postcode{201210}, \country{P. R. China}}}

\affil[13]{\orgdiv{National Key Laboratory of Computational Physics},\linebreak \orgname{
Institute of Applied Physics and Computational Mathematics}, \orgaddress{\city{Beijing}, \postcode{100094}, \country{P. R. China}}}

\affil[14]{\orgname{HEDPS, CAPT, College of Engineering}, \orgname{Peking University}, \orgaddress{\city{Beijing}, \postcode{100871}, \country{P. R. China}}}

\affil[15]{\orgname{National Supercomputing Center in Chengdu}, \linebreak \orgaddress{\city{Chengdu}, \postcode{610000}, \country{P. R. China}}}


\date{\today}

\abstract{Controlling magnetic phases and properties through composition and external conditions is central to the development of magnetic materials. Atomistic foundation models offer a route to predicting such behaviour across diverse materials, but magnetic materials remain challenging because their atomic structure and local moments are strongly coupled, while temperature and disorder drive magnetic moments into non-collinear states with evolving orientations and magnitudes. Broadly pretrained atomistic models still sample only limited magnetic diversity, while finite-temperature descriptions of coupled lattice and magnetic fluctuations remain largely material-specific. Here we introduce MagPT, pretrained across 48 elements on 912,492 spin–lattice configurations spanning structural, orientational and longitudinal magnetic fluctuations over 0 $\sim$ 1,000 K and 0 $\sim$ 10 GPa. By treating species, atomic positions and vector moments together, MagPT distinguishes magnetic states that share the same atomic structure but differ in spin configuration and energy. It captures magnetic behaviour across composition, temperature and length scales, from Slater–Pauling-type composition–magnetism relations in transition-metal alloys and finite-temperature phase evolution in elemental ferromagnets and magnetic shape-memory Heusler alloys, to emergent long-wavelength non-collinear spin textures in chiral magnets and local chemistry–magnetism coupling in high-entropy alloys. MagPT moves atomistic studies of magnetic materials beyond material-specific approaches, establishing a general framework for exploring and designing magnetic matter across diverse chemistries and conditions.}

\keywords{Magnetic materials, Atomistic foundation models, Finite-temperature magnetism, Composition–structure–property relationships}

\maketitle

\newpage

Tailoring magnetic properties through composition and structural control requires understanding how local atomic environments shape both local moments and collective magnetic order. At fixed composition, lattice strain can reorganize non-collinear textures, as illustrated by pronounced skyrmion deformations in FeGe \cite{Shibata2015SkyrmionStrain}. Temperature-driven magnetic transformations can additionally involve atom movement and changes in local-moment magnitude, as in the antiferromagnetic-to-ferromagnetic transition of FeRh \cite{Cherifi2014FeRh,Lee2025LongitudinalSpinPumping}. In concentrated alloys, chemical complexity introduces a further dependence on both elemental composition and local atomic arrangement, linking the bulk magnetic response to a distribution of site-specific magnetic environments \cite{Kormann2015HEA,Walsh2021MagneticSRO,Niu2018HEA}. Across these systems, predicting magnetic behaviour requires resolving how structure and chemistry alter the competition between magnetic configurations. Even for the same atomic arrangement, different relative orientations and magnitudes of local moments can have different energies \cite{Ma2015ConstrainedDFT}. A common atomistic description of these coupled responses must therefore retain atomic species \(Z_i\), positions \(\mathbf r_i\) and vector local moments \(\mathbf m_i\) within an energy landscape \(E\{Z_i,\mathbf r_i,\mathbf m_i\}\), providing a microscopic basis for composition-dependent magnetism, non-collinear order and finite-temperature response.

The central challenge is to connect this coupled energy landscape to experimentally observable magnetic behaviour across materials, predicting how composition, structure and temperature alter the response and identifying the microscopic origins of these changes. First-principles methods resolve magnetic energetics but remain costly for exploring chemically heterogeneous and thermally fluctuating states, whereas scalable magnetic models often require material-specific interactions or parameters \cite{Ma2015ConstrainedDFT,Abert2019Micromagnetics,Landau1974MagneticMC,Skubic2008ASD,Hellsvik2019SLD,Chapman2022MSLP}. Atomistic foundation models address the need for reuse by learning broadly pretrained representations that can be applied directly or adapted with target-specific data \cite{Chen2022M3GNet,Deng2023CHGNet,Merchant2023GNoME,Batatia2025MACE,Zhang2024DPA2,Zhang2026DPA3,Takamoto2022PFP45,Takamoto2023PFP72,Xu2025SpinInformed,Ho2026mMACEarxiv}. For magnetism, however, chemical transferability must be accompanied by explicit access to different magnetic states. Many widely used foundation potentials condition energy on atomic species and positions, without retaining local moments as independently variable coordinates \cite{Chen2022M3GNet,Deng2023CHGNet,Merchant2023GNoME,Batatia2025MACE,Zhang2024DPA2,Zhang2026DPA3,Takamoto2022PFP45,Takamoto2023PFP72}. Explicit magnetic potentials resolve this limitation, but their development has largely focused on restricted chemical systems \cite{Novikov2022mMTPs,Rinaldi2024magneticACE,Yang2024DeepSPIN,Yu2024SpinGNNpp,Yuan2024MagNet,Gao2026STEParxiv}. Chemically broader magnetic models now provide shared pretrained representations, although their large-scale pretraining has mainly used collinear, magnetically relaxed reference states, even when the architecture supports non-collinear magnetism \cite{Xu2025SpinInformed,Ho2026mMACEarxiv}. The outstanding challenge is therefore to combine chemical breadth with systematic coverage of structural fluctuations and independently variable moment orientations and magnitudes within a common pretrained energy landscape.

Here we introduce MagPT, a magnetic atomistic foundation model that combines chemically broad pretraining with explicit learning of coupled structural and magnetic states. Active learning explores structural fluctuations, non-collinear orientations and local-moment magnitude variations, while constrained-spin DFT retains the sampled magnetic configurations during labelling. Pretraining on 912,492 configurations spanning 48 elements yields a shared energy landscape over atomic species, positions and vector moments. This single pretrained model supports direct predictions and provides a common starting point for data-efficient fine-tuning and distillation into compact specialists. Spin ablation exposes magnetic energy differences lost in a structure-only representation, while zero-shot Slater--Pauling predictions demonstrate transfer across disordered alloy compositions. Few-shot adaptation reduces target DFT requirements, and distilled specialists enable finite-temperature simulations and generate long-wavelength non-collinear textures in FeGe. In a five-component high-entropy alloy, a MagPT-derived specialist captures the experimental saturation-magnetisation trend and resolves chemical short-range order alongside environment-dependent local moments.

\section{Results}

\begin{figure}[p]
  \centering
  \includegraphics[width=\linewidth]{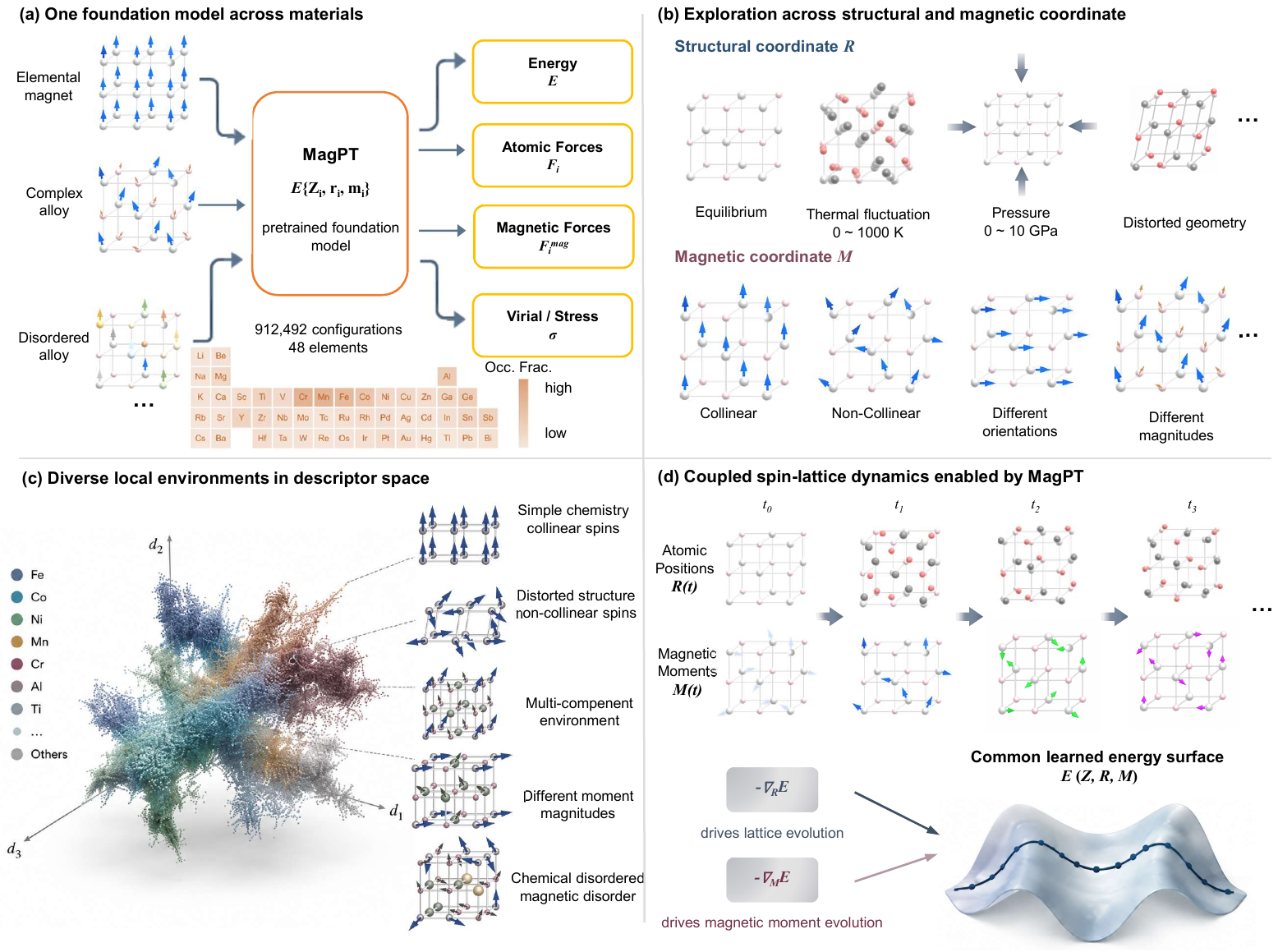}
  \caption{\textbf{MagPT is a pretrained foundation model for coupled spin--lattice states.}
  \textbf{a,} A shared model $E\{Z_i,\mathbf{r}_i,\mathbf{m}_i\}$ predicts energies, atomic forces, magnetic forces and virial/stress for elemental magnets, intermetallics, complex alloys and disordered alloys, trained on 912,492 configurations and 48 elements.
  \textbf{b,} Sampling spans equilibrium, thermal and pressure-driven structures, plus collinear/non-collinear magnetic states with varied orientations and magnitudes.
  \textbf{c,} Local environments project onto descriptor axes and are colored by element, highlighting chemical and magnetic diversity.
  \textbf{d,} MagPT-driven coupled spin--lattice dynamics: atomic positions $\mathbf{R}(t)$ and magnetic moments $\mathbf{M}(t)$ evolve simultaneously on a shared energy surface $E$, driven by $-\nabla_{\mathbf{R}}E$ and $-\nabla_{\mathbf{M}}E$.}
  \label{fig:model_overview}
\end{figure}

\subsection{MagPT couples broad chemistry to non-equilibrium spin--lattice states}

MagPT builds on the Deep Potential Attention 3 (DPA3) architecture, which employs line-graph message passing to capture many-body atomic interactions within a scalable, shared energy model \cite{Zhang2026DPA3}. By incorporating local vector magnetic moments as explicit degrees of freedom, MagPT extends this architecture to learn a unified spin--lattice energy landscape that couples chemical, structural and magnetic states \cite{Yang2024DeepSPIN}. This shared model is pretrained on 912,492 labelled spin--lattice configurations involving 48 elements and a broad range of predominantly metallic and intermetallic chemistries (Fig.~\ref{fig:model_overview}a). Each configuration is represented by
\begin{equation}
\mathcal{C} = \left\{Z_i,\mathbf{r}_i,\mathbf{m}_i \right\}_{i=1}^{N},
\end{equation}
where \(Z_i\), \(\mathbf{r}_i\) and \(\mathbf{m}_i\) denote the chemical species, atomic position and local vector magnetic moment of atom \(i\), respectively. MagPT maps this joint state to a single extensive energy \(E(\mathcal{C})\), without requiring a priori decomposition into separately parameterized lattice, exchange and magnetoelastic contributions. Derivatives of this energy with respect to atomic coordinates, local magnetic moments and cell deformation provide the atomic forces, magnetic forces and virials, respectively. The same learned energy surface can therefore support structural and magnetic relaxation as well as their coupled finite-temperature evolution across different chemical environments. By sharing one pretrained representation across materials and physical tasks, MagPT is designed to provide a reusable spin--lattice energy prior rather than requiring a separately parameterized magnetic potential for each material.

The breadth required for such a foundation model extends along two complementary directions: across chemically distinct systems and across the non-equilibrium structural and magnetic states accessible within each system. Beyond the chemical diversity shown in Fig.~\ref{fig:model_overview}a, model-driven exploration with LAMMPS-TSPIN\cite{Thompson2022LAMMPS,Huang2026TSPINarxiv} expands the pretraining distribution along both structural and magnetic coordinates. Along the structural coordinates, the sampled configurations include thermally and mechanically perturbed states under conditions spanning 0--1000 K and 0--10 GPa. Along the magnetic coordinates, the dataset contains competing collinear and non-collinear moment arrangements together with longitudinal variations in moment magnitude. During exploration, an ensemble of provisional models monitors uncertainties in the atomic-force and magnetic-force channels separately, and configurations that are uncertain in either channel are labelled by constrained-spin DFT using DeltaSpin in ABACUS and returned to training\cite{Zhou2025ABACUS,Cai2023DeltaSpin} (Fig.~\ref{fig:model_overview}b). Atomic-force uncertainty identifies underrepresented structural environments, whereas magnetic-force uncertainty identifies poorly represented moment orientations and magnitudes, including magnetic states associated with otherwise similar atomic geometries. The reference labels are obtained from non-collinear DFT with spin--orbit coupling and constrained local-moment vectors, so that the prescribed magnetic state is retained during electronic self-consistency rather than replaced by an unconstrained magnetic state. The descriptor projection in Fig.~\ref{fig:model_overview}c provides a qualitative view of the resulting local-environment diversity, while representative configurations illustrate that chemical identity, atomic geometry, moment orientation and moment magnitude can vary simultaneously within the sampled states. Importantly, this expansion of the pretraining distribution does not come at the expense of predictive accuracy: on held-out in-distribution configurations, MagPT reaches RMSEs of approximately 11~meV atom$^{-1}$ in energy, 116~meV~\angstrom$^{-1}$ in atomic force and 35~meV~$\mu_{\mathrm B}^{-1}$ in magnetic force. The resulting corpus therefore combines cross-material chemical diversity with off-equilibrium structural and magnetic diversity within a single pretrained model, providing the state-space breadth needed for a common spin--lattice representation to be reused across different magnetic materials and physical conditions.

\subsection{Explicit local moments resolve magnetic-state aliasing in spin--lattice energy landscapes}

\begin{figure}[p]
  \centering
  \includegraphics[width=0.8\linewidth]{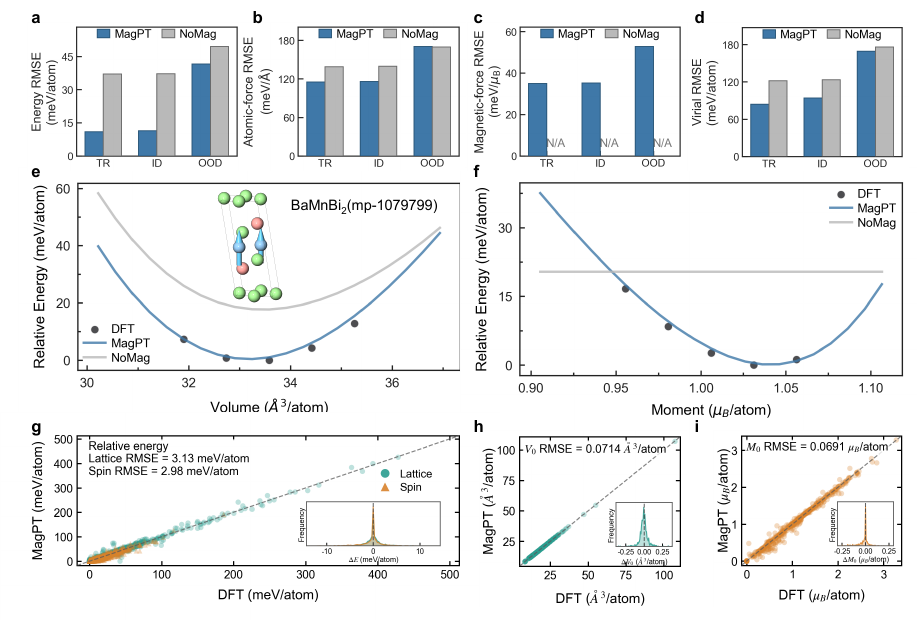}
  \caption{\textbf{Explicit local moments resolve magnetic-state aliasing.}
  \textbf{a--d,} RMSEs in energy, atomic force, magnetic force and virial for MagPT and a \textit{spin-ablated} baseline on the training set (TR), an in-distribution test set (ID), and a formula-held-out test set (OOD). Both models use the same configurations and shared energy, atomic-force and virial targets; only MagPT receives local magnetic moments and magnetic-force supervision. Magnetic force is undefined for the \textit{spin-ablated} model.
  \textbf{e,f,} Energy landscapes of BaMnBi$_2$ (mp-1079799), one formula randomly selected from the formula-OOD set. All energies are referenced to the DFT minimum.
  \textbf{e,} Relative energies under isotropic volume scaling. MagPT follows the low-energy DFT branch, whereas the \textit{spin-ablated} model predicts an elevated energy profile.
  \textbf{f,} Relative energies under uniform scaling of local magnetic moments at fixed geometry. MagPT captures the longitudinal DFT magnetic-energy branch; the \textit{spin-ablated} prediction remains constant because its inputs do not change.
  \textbf{g,} MagPT vs.\ DFT relative energies for lattice- and moment-scaling scans across all systems. Insets: residual distributions.
  \textbf{h,i,} Predicted equilibrium volume $V_0$ and minimum-energy average moment $M_0$ versus DFT. Insets: residual distributions.}
  \label{fig:benchmark}
\end{figure}

Having established a joint \(Z\)--\(R\)--\(M\) pretraining space, we next asked a more fundamental question: whether \(M\) carries state information that cannot be represented by \(Z\) and \(R\) alone.  Within the constrained-spin DFT reference used here, identical chemical species and atomic coordinates can support distinct local-moment configurations with different energies. Projecting such states onto a structure-only representation therefore maps physically distinct magnetic states onto the same model input, so that the projected structure--energy relation is no longer single-valued. A deterministic structure--only potential must therefore collapse distinct magnetic branches onto a common representation. To test this representational ambiguity, we compared MagPT with a \textit{spin-ablated} baseline trained on the same atomic configurations and energy, atomic-force and virial targets, but without local-moment inputs and, consequently, without magnetic-force supervision. Removing the magnetic variables increases the energy RMSE from approximately 11 to 37~meV atom\(^{-1}\) on both the training and ID sets, accompanied by smaller but systematic degradations in atomic-force and virial accuracy (Fig.~\ref{fig:benchmark}a,b,d). Because the degradation is already present on the training data, it cannot be explained by extrapolation alone and is consistent with information discarded by the structure-only projection. Magnetic-force errors are reported only for MagPT, because this quantity is undefined for a representation without local-moment degrees of freedom. (Fig.~\ref{fig:benchmark}c).

We first tested this ambiguity directly by varying only the prescribed local moments while keeping the chemical species and atomic geometry fixed. For this direct test outside the training formulae, we selected BaMnBi$_2$, one formula randomly drawn from the formula-out-of-distribution(OOD) set. Constrained-spin DFT exhibits a non-trivial energy variation as the prescribed local moments are uniformly scaled, which MagPT reproduces closely (Fig.~\ref{fig:benchmark}f). By construction, the \textit{spin-ablated} model must assign the same energy to every point because \(Z\) and \(R\) remain unchanged. This fixed-geometry scan therefore directly exposes magnetic-state aliasing in a structure-only representation. Omitting \(M\) also prevents a structure-only model from resolving the appropriate magnetic branch as the lattice is deformed. Under isotropic volume scaling with the prescribed magnetic configuration retained, MagPT follows the corresponding DFT energy branch, whereas the \textit{spin-ablated} model yields a different profile (Fig.~\ref{fig:benchmark}e), consistent with distinct magnetic branches being collapsed onto a common structural representation.

To determine whether this branch-resolving capability extends beyond a single example, we evaluated the pooled collection of lattice- and moment-scaling scans. MagPT reproduces relative energies along the lattice and magnetic coordinates with RMSEs of 3.13 and 2.98~meV atom$^{-1}$, respectively (Fig.~\ref{fig:benchmark}g). The agreement extends to the locations of the minima along both coordinates: the predicted minimum-energy volumes and average local-moment magnitudes have RMSEs of 0.071~\angstrom$^3$ atom$^{-1}$ and 0.069~$\mu_{\mathrm B}$ atom$^{-1}$, respectively (Fig.~\ref{fig:benchmark}h,i). Together, these results show that the local magnetic state is not an auxiliary descriptor inferred from structure, but a non-redundant state coordinate required to define a single-valued spin--lattice energy landscape.

\subsection{Zero-shot Recovery Slater--Pauling behaviour in explicitly disordered alloys}

\begin{figure}[p]
  \centering
  \includegraphics[width=0.8\linewidth]{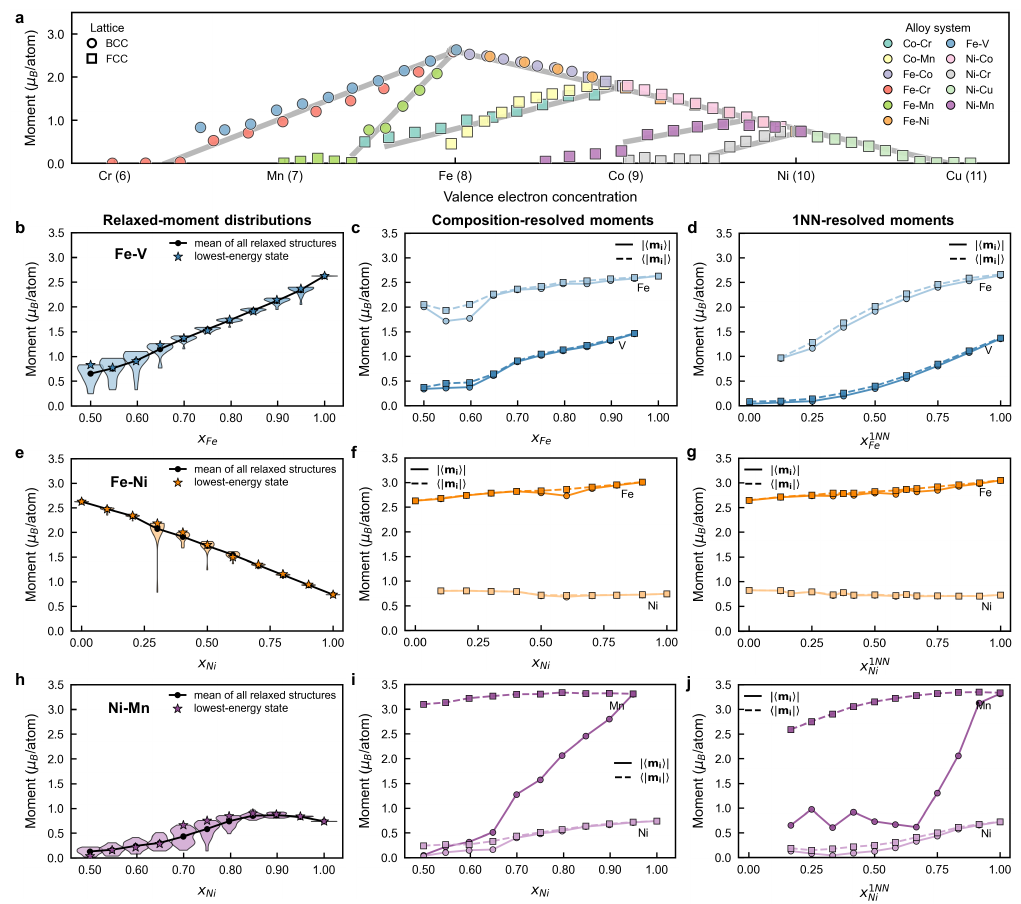}
  \caption{\textbf{Zero-shot Slater--Pauling behaviour across binary alloys.}
  \textbf{a,} Magnetic moment as a function of valence electron concentration for binary alloy systems predicted by the pretrained model without system-specific fitting.
  \textbf{b-d,} Fe--V relaxed candidate moments, element-resolved local moments and nearest-neighbour-environment-resolved moments.
  \textbf{e-g,} Corresponding analysis for Fe--Ni.
  \textbf{h-j,} Corresponding analysis for Ni--Mn, where changes in Mn non-collinearity dominate the composition- and environment-dependent magnetic response.}
  \label{fig:slater_pauling}
\end{figure}

Having established that explicit magnetic coordinates are required to resolve the spin–lattice energy landscape, we next ask whether the pretrained representation can recover emergent composition–magnetism relations without alloy-specific training(Fig.~\ref{fig:slater_pauling}). No alloy-specific fine-tuning or Slater–Pauling-targeted data are used, and the tested composition grid, 256-atom supercells, random occupational realizations and relaxation trajectories are all absent from pretraining.

For each composition, we sample competing BCC and FCC structures, magnetic initial states and independent occupational realizations, allowing the relaxed magnetic state to emerge from explicit structural, chemical and spin degrees of freedom. The resulting net moments reproduce the characteristic Slater--Pauling evolution across the tested binary alloys \cite{Williams1983SlaterPauling}. Importantly, the calculations retain the distribution arising from explicit occupational disorder, rather than replacing each composition by a single effective medium. Some compositions exhibit narrow moment distributions, indicating weak sensitivity to local occupation, whereas others show broad distributions that signal strong coupling between chemical environment and magnetic-state selection.

The composition--moment trends recovered in different regions of the Slater--Pauling curve reflect distinct microscopic contributions in the three alloy families. In Fe--V, both Fe and V moment magnitudes vary systematically with composition and nearest-neighbour occupation. In Fe--Ni, the local moments are comparatively insensitive to occupation, so the net moment is dominated by the changing elemental fractions. In Ni--Mn, large local Mn moments persist, but their projected contribution changes through non-collinear or compensating alignment. MagPT thus recovers a collective Slater--Pauling-type composition--moment relation while retaining the local chemical environments, moment magnitudes and moment orientations from which that relation emerges, indicating that pretraining has encoded transferable magnetic chemistry rather than a composition-specific empirical rule.


\subsection{Data-efficient construction of material-specific spin–lattice models}

\begin{figure}[p]
  \centering
  \includegraphics[width=0.8\linewidth]{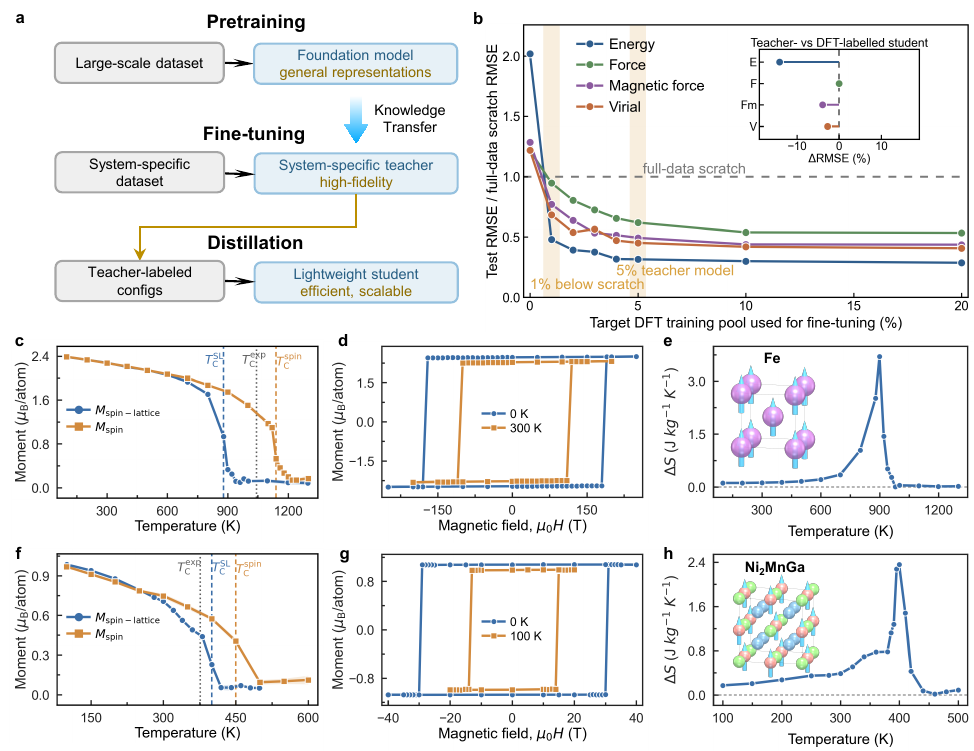}
  \caption{\textbf{Few-shot adaptation enables finite-temperature magnetic modelling from limited DFT data.}
  \textbf{a,} Workflow from pretrained MagPT to few-shot fine-tuned teacher and compact distilled specialist model.
  \textbf{b,} Data efficiency on an out-of-distribution Fe dataset.
  RMSEs (energy, forces, magnetic forces, virial) normalised to a scratch-trained baseline (dashed line).
  Shaded band: 1\% (all metrics below scratch baseline) and 5\% (teacher for distillation) data regimes.
  Inset: teacher- vs.\ DFT-supervised students; negative values favour the teacher-labelled student.
  \textbf{c,} Temperature dependence of the magnetic moment of Fe from fixed-lattice and coupled spin--lattice simulations.
  Vertical lines: experimental and simulated transition temperatures.
  \textbf{d,} Magnetisation of Fe during increasing and decreasing magnetic-field sweeps at 0 and 300~K.
  \textbf{e,} Isothermal magnetic entropy change of Fe for a magnetic-field change of \(\Delta\mu_0 H\)=2~T.
  \textbf{f--h,} Same as \textbf{c--e}, but for Ni\(_2\)MnGa.}
  \label{fig:specialization}
\end{figure}

MagPT substantially reduces the additional constrained-spin DFT data required to construct a material-specific spin--lattice model. On a separate Fe dataset not included in pretraining, fine-tuning on only 1\% of the available target data already yields lower test errors than a model of the same architecture trained from random initialisation on the complete target dataset, while most of the attainable improvement is reached with 5\% (Fig.~\ref{fig:specialization}b). Material-specific accuracy can therefore be reached by adapting a pretrained magnetic representation rather than learning the coupled spin--lattice energy landscape anew. 

The resulting fine-tuned model can also serve as a teacher for compact spin--lattice models, reducing the need for direct DFT labelling during student training. We therefore used the 5\% fine-tuned MagPT model to label configurations generated during further exploration of the target configuration space and trained a compact student model on the expanded dataset. The resulting student achieves comparable or even slightly lower errors (-14~\%) across all four quantities than a student of the same architecture trained on DFT labels, as evaluated on the common held-out DFT test set. Within the tested regime, compact-model construction can therefore rely on a pretrained teacher adapted with limited target DFT data, rather than on direct DFT labelling of the full student-training set.

The compact students enable efficient simulations of temperature- and field-dependent magnetism. Relative to fine-tuned MagPT, the student models accelerate inference by two to three orders of magnitude, enabling substantially larger spin--lattice simulations while retaining finite-temperature magnetic behaviour. In Fe and Ni$_2$MnGa, the distilled models resolve the temperature evolution of magnetic order and the associated magnetic transitions (Fig.~\ref{fig:specialization}c,f). Allowing the atomic positions to evolve shifts these transitions relative to fixed-lattice spin simulations, showing directly that lattice dynamics modifies the finite-temperature magnetic response under the present simulation conditions. The same models also capture field-driven magnetisation reversal (Fig.~\ref{fig:specialization}d,g) and the associated isothermal magnetic entropy changes (Fig.~\ref{fig:specialization}e,h), extending the finetuning-to-distillation workflow from local energetic accuracy to finite-temperature and field-dependent magnetic behaviour.

\subsection{Emergent long-wavelength chiral spin textures in FeGe}

\begin{figure}[p]
  \centering
  \includegraphics[width=1.0\linewidth]{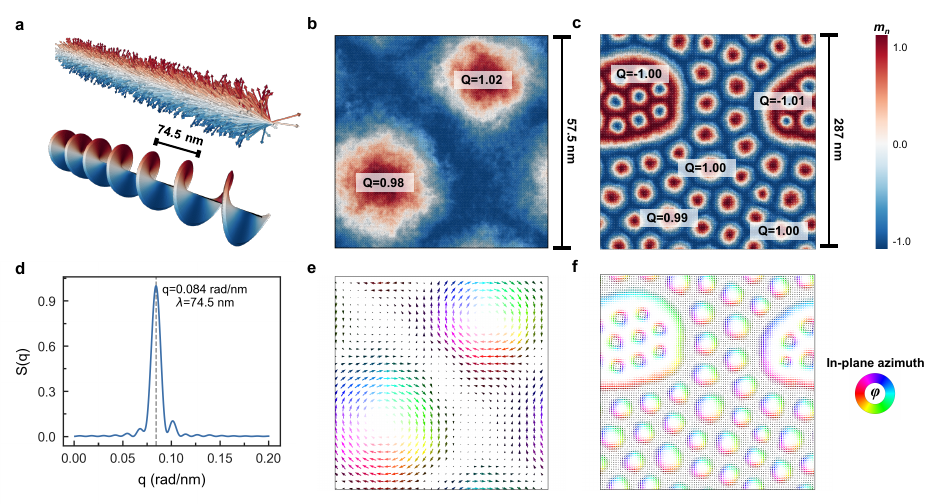}
  \caption{\textbf{Emergent long-wavelength chiral spin textures in FeGe.}
  \textbf{a-c,} Spatial distributions of the normalized surface-normal magnetic-moment component, $m_n$, for a helical state and skyrmion textures. 
  The helical texture in (\textbf{a}) has a periodicity of 74.5~nm, whereas the skyrmion textures in (\textbf{b,c}) exhibit alternating regions of positive and negative $m_n$, with the corresponding topological charge $Q$ evaluated for representative skyrmion regions indicated. 
  \textbf{d,} Spin structure factor $S(q)$ of the helical texture in (\textbf{a}), showing a well-defined peak at $q=0.084$~rad~nm$^{-1}$, corresponding to $\lambda=74.5$~nm. 
  \textbf{e,f,} In-plane azimuthal angle $\varphi$ for the skyrmion textures in (\textbf{b,c}), respectively, highlighting the characteristic winding of the in-plane magnetization around the skyrmion cores.
  The colour scales represent $m_n$ in (\textbf{a-c}) and the in-plane azimuthal angle $\varphi$ in (\textbf{e,f}).}
  \label{fig:fege}
\end{figure}

Having established a data-efficient route to compact models for finite-temperature magnetism, we next ask whether the learned orientational energy landscape is sufficient to recover collective non-collinear order over length scales far beyond those directly accessible to constrained-spin DFT.  B20-FeGe provides a stringent test because its helical and skyrmionic states arise from the delicate competition between exchange, Dzyaloshinskii--Moriya interactions and magnetic anisotropy, producing magnetic textures whose characteristic length scales extend over tens of nanometres\cite{Yu2011FeGe,Yang2024FeGeSkyrmionBags,Koretsune2015DMI,Burn2020FeGe}. Such states are particularly demanding for a \(Z\)--\(R\)--\(M\) potential: nearly identical atomic structures can host spatially distinct magnetic configurations, while small local energetic differences must remain coherent over hundreds of atomic spacings to select the resulting long-wavelength texture. To test whether these textures emerge from the learned magnetic energy landscape rather than from an imposed magnetic template, we initialize all three large-cell calculations with random local-moment orientations, without seeding either helical or skyrmionic order (Fig.~\ref{fig:fege}). The largest system spans \(287\times287\) nm\(^2\) and contains 2,976,800 atoms, while no DFT configuration in the training data contains more than 60 atoms, extending the atomistic magnetic description to \(5\times10^4\) times the size of the cells used for first-principles training. 

Starting from the disordered magnetic state, the elongated FeGe cell spontaneously develops a coherent helix with an estimated wavelength of 74.5 nm. The emergent wavelength, obtained from the dominant peak in the spin structure factor S(q), is close to the experimentally established characteristic period of approximately 70 nm in FeGe, despite no long-wavelength magnetic periodicity being imposed during texture initialization \cite{Yu2011FeGe,Grytsiuk2019FeGe}.

We next increase the dimensional and topological complexity of the test in laterally extended FeGe cells. Under an applied field of 0.1~T, a \(57.5\times57.5\) nm$^2$ cell initialized with disordered moments relaxed into two isolated skyrmionic textures, with an integrated topological charge of $Q=2$. Increasing the lateral dimensions to \(287\times287\) nm$^2$ produced a substantially more complex magnetic state containing 49 skyrmionic objects, including a skyrmion bag enclosing 12 oppositely charged ($Q=-1$) skyrmions. Integration of the topological-charge density over the full 2,976,800 atoms yields a net topological charge of $Q=25$.

Our aim here is not to reconstruct the full temperature--field phase diagram of FeGe, but to test whether locally learned first-principles magnetic energetics can recover long-wavelength and topologically non-trivial order at experimentally relevant length scales. Together, these results show that the local spin--lattice energetics learned from constrained-spin DFT remain sufficiently coherent after fine-tuning and distillation to generate collective magnetic organization across hundreds of nanometres, spanning both the characteristic helical wavelength and non-trivial magnetic topology.

\subsection{Quantitative magnetic prediction and chemical-state resolution in a high-entropy alloy}

\begin{figure}[p]
  \centering
  \includegraphics[width=\linewidth]{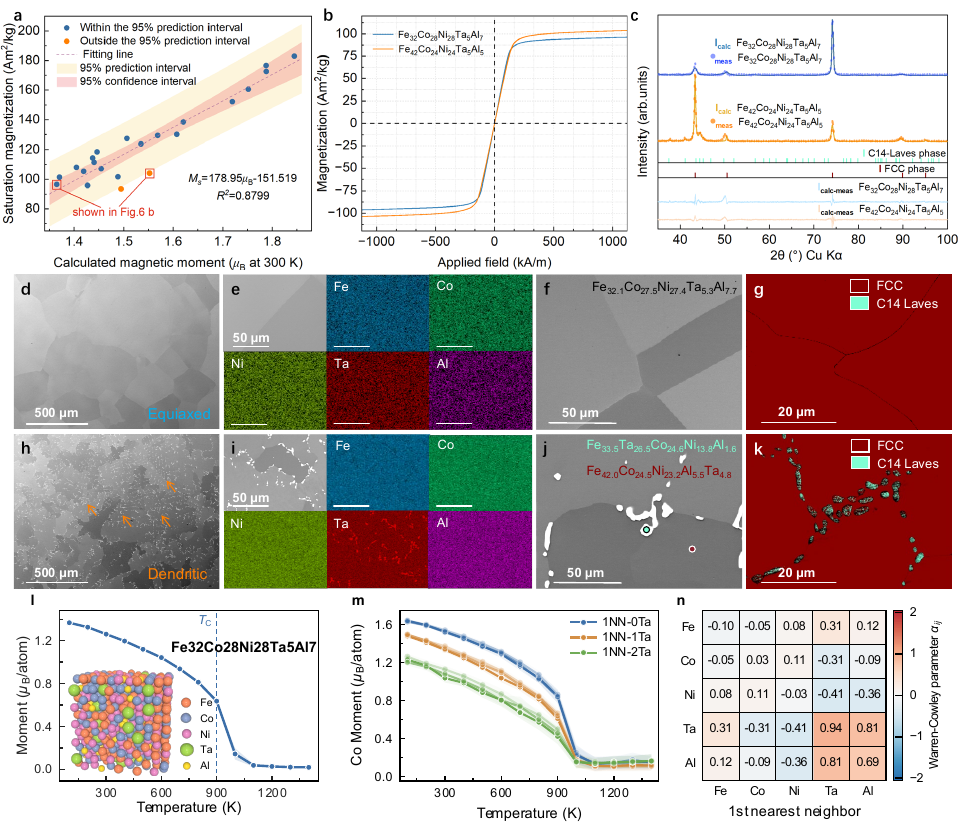}
  \caption{\textbf{Quantitative magnetic prediction and chemical-state resolution in a high-entropy alloy}
  \textbf{a,} Linear regression between experimentally measured saturation magnetization ($M_S$) and calculated magnetic moment at 300 K for 22 FeCoNiTaAl high-entropy alloys (HEAs) with different compositions.
  \textbf{b,} Room-temperature magnetic hysteresis loops of Fe$_{42}$Co$_{24}$Ni$_{24}$Ta$_5$Al$_5$ and Fe$_{32}$Co$_{28}$Ni$_{28}$Ta$_5$Al$_7$.
  \textbf{c,} Measured and Pawley-refined X-ray diffraction patterns.
  \textbf{d,h,} Backscattered-electron (BSE) micrographs, \textbf{e,i,} Energy-dispersive X-ray spectroscopy (EDS) elemental maps, \textbf{f,j,} BSE micrographs with corresponding EDS analyses, and \textbf{g,k,} Electron Back Scatter Diffraction (EBSD) phase maps of Fe$_{32}$Co$_{28}$Ni$_{28}$Ta$_5$Al$_7$ and Fe$_{42}$Co$_{24}$Ni$_{24}$Ta$_5$Al$_5$, respectively.
  \textbf{l,} Temperature-dependent mean magnetic moment from MagPT (inset: fragment of the supercell; dashed line: simulated $T_C$).
  \textbf{m,} Co moment versus temperature conditioned on 0, 1, or 2 Ta atoms in the first-neighbor shell.
  \textbf{n,} First-nearest-neighbor Warren--Cowley parameter matrix at 1000~K (negative/positive: pair enrichment/depletion).}
  \label{fig:hea}
\end{figure}

Having tested large-cell non-collinear textures in FeGe, we next evaluated a MagPT-derived specialist model against measurements for 22 experimentally synthesised alloys in the Fe--Co--Ni--Ta--Al system. Experimental saturation magnetisation, $M_S$, exhibits an approximately linear correlation with the magnetic moments calculated at 300~K ($R^2=0.8799$; Fig.~\ref{fig:hea}a), with two alloys departing from the common trend. To examine whether the experimental specimens and atomistic calculations represent comparable material states, we characterised Fe$_{32}$Co$_{28}$Ni$_{28}$Ta$_5$Al$_7$, which follows the trend, and Fe$_{42}$Co$_{24}$Ni$_{24}$Ta$_5$Al$_5$, one of the two deviating alloys.

Structural and chemical characterisation distinguishes the phase states of these two alloys. In Fe$_{32}$Co$_{28}$Ni$_{28}$Ta$_5$Al$_7$, X-ray diffraction and electron backscatter diffraction identify an FCC structure with no secondary phase detected within the sensitivity of the measurements. Elemental mapping further shows spatially uniform distributions at the measurement scale, with the average composition close to the nominal value (Fig.~\ref{fig:hea}c--g). By contrast, Fe$_{42}$Co$_{24}$Ni$_{24}$Ta$_5$Al$_5$ contains an FCC matrix and an interdendritic C14 Laves phase with pronounced chemical partitioning (Fig.~\ref{fig:hea}c,h--k). Representative EDS point analyses give approximately 26.5~at.\% Ta in the secondary phase compared with 4.8~at.\% in the matrix. Its measured $M_S$ therefore combines the responses of chemically distinct regions, whereas the corresponding calculation represents a homogeneous FCC phase at the nominal composition. This phase-state and chemical mismatch is consistent with the observed deviation, although its quantitative contribution remains unresolved. The alloy without a detected secondary phase provides a more direct reference for examining the magnetic response within the FCC state.

At this reference composition, we sampled the finite-temperature FCC alloy by alternating spin--lattice dynamics with composition-preserving element-swap Monte Carlo moves. This procedure jointly samples atomic displacements, chemical occupations and changes in both the orientations and magnitudes of local magnetic moments on the same learned energy landscape. Non-collinear magnetic configurations and longitudinal moment fluctuations are therefore retained alongside lattice motion and chemical rearrangements. The calculated magnetisation decreases continuously with temperature and approaches zero near the simulated Curie temperature (Fig.~\ref{fig:hea}l). These calculations extend the room-temperature comparison to a predicted finite-temperature response while retaining the site-resolved variables needed to examine its microscopic origin.

The sampled configurations reveal local chemical correlations that are not specified by the overall alloy composition. At 1000~K, the first-neighbour Warren--Cowley matrix shows an enrichment of Ni--Ta, Ni--Al and Co--Ta contacts relative to random occupation (Fig.~\ref{fig:hea}n). To examine the association between chemical coordination and local magnetism, we grouped Co atoms by the number of Ta atoms in their first coordination shell. Throughout the magnetically ordered regime, the mean Co moment decreases systematically from sites with no Ta nearest neighbours to those with one and two Ta neighbours (Fig.~\ref{fig:hea}m). Thus, even within the same FCC alloy, atoms of the same magnetic element exhibit distinct mean moments depending on their local chemical environments. Together, the pair statistics and coordination-conditioned moments resolve chemical variations in local magnetism that are hidden in the bulk magnetic observable.

Previous calculations have established the sensitivity of alloy magnetism to local environments and chemical short-range order \cite{Schneeweiss2017CrMnFeCoNi,Feng2017SRO,Woodgate2023MagneticSRO,Su2024SpinCE,Chen2021CSRO}, while material-specific magnetic machine-learning potentials have enabled explicit finite-temperature spin--lattice simulations \cite{Chapman2022MSLP,Lavrentiev2016MCE}. Here, a single specialist derived from chemically broad magnetic pretraining connects composition-dependent experimental trends with temperature-dependent magnetic order and environment-resolved atomic moments. The HEA results therefore extend the transferred representation beyond a composition--property relation, providing an atomistic basis for interpreting magnetic behaviour through the coupled chemical, structural and vector-magnetic states of a complex alloy.

\section{Discussion}

MagPT shows that magnetic materials are more naturally described on a coupled chemical–structural–magnetic energy landscape than on a structural surface with magnetism treated as an auxiliary response. By retaining vector local moments as explicit state coordinates, MagPT connects experimentally controlled variables—composition, strain and temperature—with the local chemical, structural and magnetic states that generate the observed response. This is particularly important when nominally similar materials develop different magnetic behaviour because processing changes local order, lattice state or magnetic configurations.

The significance of MagPT as a foundation model is therefore not simply reduced target-data requirements. Its broader value is to replace material-by-material construction of magnetic models with adaptation from a shared magnetic prior, enabling a continuous workflow from broad exploration across compositions and conditions to quantitative refinement and large-scale finite-temperature simulation.

For materials research, MagPT also creates a route from macroscopic magnetic properties back to their microscopic origin. In compositionally complex alloys, measured magnetisation reflects not only nominal composition, but also phase constitution, chemical partitioning, short-range order and environment-dependent local moments. Extending MagPT towards phase-resolved microstructures, interfaces and chemically heterogeneous states could therefore connect composition and processing, through local atomic environments, to experimentally observable magnetic behaviour.

\section{Methods}

\subsection{Foundation-dataset construction and active learning}

The MagPT dataset is generated with a DP-GEN2 active-learning workflow\cite{Zhang2020dpgen}. Initial structures are selected from magnetic entries in the Materials Project\cite{Horton2025MaterialsProject}. To keep first-principles labelling tractable while retaining chemical diversity, the present dataset uses systems with no more than four elements and no more than 60 atoms in the primitive cell; nonmetal and rare-earth elements are excluded in this release because of Hubbard-\(U\) and localized \(f\)-electron complications. At each active-learning iteration, an ensemble of provisional magnetic models is trained, used to explore spin--lattice configurations, and then updated with newly labelled configurations. Model-driven exploration uses TSPIN-based\cite{Huang2026TSPINarxiv} spin--lattice dynamics over 0--1000 K and 0--10 GPa, with variations in compositions, atomic positions and magnetic initial states. Configurations with large ensemble uncertainty in atomic or magnetic forces are selected for DFT labelling.

\subsection{Constrained-spin DFT labelling}

Selected configurations were labelled with ABACUS using the \textsc{DeltaSpin} constrained-spin formalism to stabilise non-ground-state magnetic configurations\cite{Zhou2025ABACUS,Cai2023DeltaSpin}. Calculations used a plane-wave basis, fully relativistic optimized norm-conserving Vanderbilt pseudopotentials and the PBE exchange-correlation functional. The kinetic-energy cutoff was 100 Ry, and Brillouin-zone sampling used a reciprocal-space spacing of 0.14 Bohr$^{-1}$. Electronic occupations were treated with Gaussian smearing of 0.01 Ry. Energy and charge convergence thresholds were set to \(10^{-5}\), the magnetic-constraint convergence threshold was set to \(10^{-7}\), and crystal symmetry was disabled. All training labels were generated from non-collinear DFT calculations with spin–orbit coupling enabled, while DeltaSpin constrained the local-moment vectors during self-consistency. The retained labels comprise total energies, atomic forces, magnetic forces and virials.

\subsection{MagPT architecture and pretraining}

MagPT is trained with the DeePMD-kit/DeepSpin framework using atomic species, positions and local magnetic moments as input variables\cite{Zhang2025DeepMDkitv3,Yang2024DeepSPIN}. Atomic environments are encoded with the DPA3 descriptor, using a 6.0 \angstrom cutoff for environmental embeddings and a 4.0 \angstrom cutoff for angular features\cite{Zhang2026DPA3}. The representation network contains six interaction layers with embedding dimensions of 256 for atomic channels and 128 for edge channels, and the fitting network contains three fully connected layers with 240 neurons per layer and residual connections. Training uses float32 arithmetic, an exponentially decaying learning rate from \(10^{-3}\) to \(10^{-5}\), and 1,000,000 optimization steps with automatically adjusted batch sizes. The loss jointly optimizes energy, atomic force, magnetic force and virial targets. Ensembles of independently trained models are used during active learning to estimate uncertainty. The pretrained model is used directly for zero-shot slater-pauling calculations and also provides the starting point for subsequent few-shot fine-tuning and teacher-student distillation. 

\subsection{Fine-tuning and distillation}

For each target material, the pretrained MagPT teacher was fine-tuned on the corresponding DFT-labelled dataset for 50,000 steps. All architectural hyperparameters were unchanged from pretraining stage, including the 6.0~\angstrom environmental cutoff, the 4.0~\angstrom angular cutoff, six DPA3 interaction layers, the atomic and edge embedding dimensions, and the three-layer fitting network. Only the learning-rate schedule was restarted with the same exponential decay parameters used during pretraining. The resulting target-specific fine-tuned MagPT model served as the teacher for subsequent dataset expansion and distillation. 

Fe and Ni$_2$MnGa were distilled into separate magnetic Deep Potential models implemented with the TensorFlow backend of DeePMD-kit. The DP model retained magnetic degrees of freedom, while using substantially reduced network capacities compared to the MagPT teacher. Specifically, these models employed a DeepPot-SE descriptor (se\_e2\_a) with a cutoff radius of 6.0 \angstrom (smoothed from 4.8 \angstrom), an embedding network of [25, 50, 100] neurons, and a fitting network of [240, 240, 240] neurons. Training proceeded for 400,000 steps with an exponentially decaying learning rate from \(10^{-3}\) to \(3.51 \times 10^{-8}\).

The larger FeGe and high-entropy-alloy simulations used separately distilled NEP-spin models implemented as a spin-enabled extension of the neuroevolution-potential framework \cite{Fan2021NEP} following C. Chen et al (in preparation). NEP-spin augments the standard structural descriptor with local magnetic channels and represents the coupled energy as a sum of local contributions depending on both atomic and magnetic environments; atomic forces, magnetic forces and virials are obtained from derivatives of this common energy. The radial and angular structural cutoffs were 6.0 and 5.0~\angstrom, respectively, and the magnetic cutoff was 6.0~\angstrom. Optimisation used batches of 1,000 configurations for 100,000 generations.

\subsection{Application to spin–lattice dynamics}

Finite-temperature and field-dependent simulations were performed in LAMMPS using the TSPIN implementation \cite{Thompson2022LAMMPS,Huang2026TSPINarxiv}, which synchronously propagates atomic positions and local magnetic moments under the atomic and magnetic forces derived from the same model energy. Static relaxations were performed with full relaxation of atomic positions, cell vectors and magnetic moments. Coupled finite-temperature spin-lattice trajectories were generated in the NPT ensemble in the NPT ensemble, with external magnetic fields introduced through the Zeeman contribution. For the high-entropy alloy, spin--lattice dynamics was further coupled to chemical Monte Carlo sampling at fixed global composition. Every 100 dynamics steps, 100 element-swap attempts were performed and accepted according to the Metropolis criterion, allowing the structural, magnetic, and chemical degrees of freedom to evolve together.

Physical properties were subsequently extracted from the equilibrated trajectories. Temperature-dependent magnetic moments were averaged over at least 10,000 steps after equilibration, and Curie temperatures were identified from the maximum slope of the magnetisation curve. Magnetic entropy changes were obtained by integrating the Maxwell relation $\Delta S = \int_0^H (\partial M/\partial T)_H dH$ \cite{Oliveira2010MagnetoCaloric}. For FeGe, the helical wavelength was extracted from the dominant peak of the spin structure factor, 
$$
S(\mathbf{q})=\frac{1}{N}\left\langle\left|\sum_j \hat{\mathbf{S}}_j e^{i\mathbf{q}\cdot\mathbf{R}_j}\right|^2\right\rangle,
$$
where $\hat{\mathbf{S}}_j = \mathbf{S}_j/|\mathbf{S}_j|$, with the helical period given by $\lambda=2\pi/q_{\max}$. Topological skyrmion textures were identified and quantified using the topological charge,
$$
Q=\frac{1}{4\pi}\int \hat{\mathbf{m}}\cdot
\left(\partial_x\hat{\mathbf{m}}\times\partial_y\hat{\mathbf{m}}\right)\,dx\,dy,
$$
where $\hat{\mathbf{m}}=\mathbf{M}/|\mathbf{M}|$. For the high-entropy alloy, chemical short-range order was quantified through the Warren--Cowley parameter, $\alpha_{ij}^{k}=1-P_{ij}^{k}/c_j$, where $P_{ij}^{k}$ is the probability of finding element $j$ in the $k$-th nearest-neighbour shell around element $i$, and $c_j$ is the global concentration of element $j$.

\subsection{Materials preparation}

Twenty-two Fe--Co--Ni--Ta--Al high-entropy alloys (HEAs) were prepared by arc melting under an argon atmosphere using high-purity constituent elements (purity $>$ 99.8~wt.\%). The as-cast HEAs were subsequently homogenized at \SI{1200}{\degreeCelsius} for \SI{2}{\hour} in a muffle furnace to improve compositional homogeneity throughout the ingots. The nominal compositions of all alloys are listed in Table~S2 of the Supplementary Materials.

\subsection{Microstructural characterization}

X-ray diffraction (XRD) measurements were performed using a SmartLab \SI{9}{\kilo\watt} diffractometer with Cu K$\alpha$ radiation. The XRD patterns were fitted using Pawley refinement. Backscattered electron (BSE) imaging, energy-dispersive X-ray spectroscopy (EDS), and electron backscatter diffraction (EBSD) were carried out using a ZEISS Sigma 360 field-emission scanning electron microscope. An accelerating voltage of \SI{15}{\kilo\volt} was used for BSE imaging and EDS analysis, while EBSD measurements were conducted at \SI{20}{\kilo\volt}.

\subsection{Magnetic response measurements}

Room-temperature hysteresis loops ($M$--$H$ curves) of twenty-two Fe--Co--Ni--Ta--Al HEAs were measured using a CX-VSM20T vibrating sample magnetometer (VSM). Cubic specimens of $3 \times 3 \times 1$~mm$^3$ were used for each measurement.

\section{Data availability}
The main data supporting the findings of this study are available in the Article and its Supplementary Information. The full 912492 dataset and the trained MagPT are not publicly available at present but may be released in the future. A subset of the dataset comprising 399,168 entries, together with the MagPT model retrained exclusively on these 399,168 frames, are openly available on GitHub at https://github.com/MagTheoryLab/MagPT. The released dataset is hosted on Hugging Face (https://huggingface.co/datasets/shiruosong/MagPT). The remaining data are available from the corresponding author B. Xu upon request.

\section{Code availability}
The software packages used for first-principles calculations, model training and spin--lattice simulation are cited in the Methods.

\section{Acknowledgements}
We acknowledge the support from National Natural Science Foundation of China (No.~52325204) and Science Challenge Project (No.TZ2025017). We thank MetaX for providing access to a large-scale GPU cluster and  National Supercomputing Center in Chengdu for DCU nodes support. We thank the DeepModeling community and the ABACUS development team for technical support and valuable discussions.

\section{Author contributions}
R.S. generated and curated the first-principles dataset, trained and validated the models, and performed the computational simulations and analyses. C.C. contributed to model training and validation and performed the FeGe simulations. R.S., C.C., Z.H., T.Y. and X.P. contributed to computational workflow development, implementation and testing. T.Y., H.W. and B.X. contributed the DeepSPIN methodology; D. Zhang and H.W. contributed to DPA-3 development and its integration with DeepSPIN. Z.C. developed the DeltaSpin methodology. Z.C. and D. Zheng jointly implemented it in ABACUS, with D. Zheng further optimizing and accelerating the implementation. Z.H. implemented TSPIN in LAMMPS, while Z.H., T.Y., X.P. and X.L. adapted DP-GEN2. S.Y. and X.Z. supported large-scale GPU computing with the GPU resources provided by MetaX Integrated Circuits (Shanghai) Co., Ltd., and J.W. supported DCU computing at the National Supercomputing Center in Chengdu. B.L. conducted the high-entropy-alloy experiments and contributed to data interpretation. H.W. and B.X. provided methodological guidance and supervised computational development, while L.H. supervised the experimental work. B.X. and K.W. co-led the project and guided its scientific direction. R.S., B.X. and L.H. wrote the manuscript, with contributions from B.L. All authors discussed the results and revised the manuscript.

\section{Competing interests}
The authors declare no competing interests.

\bibliography{references}
\end{document}